\documentclass[12pt]{article}
\pdfoutput=1

\usepackage{cite}
\usepackage{booktabs}
\usepackage[english]{babel}
\usepackage{amsmath,amssymb,amsbsy,amstext, amsthm, simplewick}
\usepackage{hyperref}
\usepackage{graphicx}
\usepackage{amsfonts}
\usepackage{amssymb}
\usepackage[small]{caption}
\usepackage{upgreek}
\usepackage[svgnames,dvipsnames,x11names,table]{xcolor}
\usepackage{multirow}
\usepackage{geometry}
\usepackage[hang,flushmargin]{footmisc}
\usepackage{bm}
\usepackage{braket}
\usepackage{subcaption}
\usepackage{mathtools}
\usepackage{setspace}
\usepackage{cleveref}
\usepackage{comment}
\usepackage{scalerel}
\usepackage[normalem]{ulem}
\usepackage{slashed}
\usepackage{enumitem}
\usepackage{dsfont}
\usepackage{tikz}
\usetikzlibrary{decorations.markings}
\usetikzlibrary{shapes.misc}

\makeatletter
\g@addto@macro\bfseries{\boldmath}
\makeatother

\hypersetup{
    colorlinks=true,
    linkcolor={red!50!black},
    citecolor={blue!50!black},
    urlcolor={blue!80!black}
}

\usepackage{colortbl}

\makeatletter
\newlength{\apb@width}
\newcommand{\autoparbox}[2][c]{\settowidth{\apb@width}{#2}\parbox[#1]{\apb@width}{#2}}

\makeatother

\definecolor{lightgray}{gray}{0.9}

\usepackage[framemethod=default]{mdframed}
\newmdenv[skipabove=7pt,
skipbelow=7pt,
rightline=false,
leftline=false,
topline=false,
bottomline=false,
backgroundcolor=gray!10,
linecolor=gray,
innerleftmargin=5pt,
innerrightmargin=5pt,
innertopmargin=5pt,
innerbottommargin=5pt,
leftmargin=0cm,
rightmargin=0cm,
linewidth=4pt]{eBox}

\usepackage[most]{tcolorbox}
\tcbset{colback=white, colframe=black,
        highlight math style= {enhanced, %<-- needed for the ?remember? options
            colframe=red,colback=red!10!white,boxsep=0pt}
        }
\definecolor{light-gray}{gray}{0.95}

\crefname{table}{Table}{Tables}
\crefname{equation}{Eq.}{Eqs.}
\crefname{appendix}{App.}{Apps.}
\crefname{section}{Sec.}{Secs.}
\crefname{figure}{Fig.}{Figs.}

\numberwithin{equation}{section}

\def\beq{\begin{equation}}
\def\eeq{\end{equation}}

\def\bea{\begin{eqnarray}}
\def\eea{\end{eqnarray}}

\def\d{{\rm d}}

\def\beq{\begin{equation}}
\def\eeq{\end{equation}}
\def\bea{\begin{eqnarray}}
\def\eea{\end{eqnarray}}

\def\d{{\rm d}}

\def\O{{\cal O}}

\def\P{{{\mathbb P}}}

\def\fnl{f_{\rm NL}}
\def\tnl{\tau_{\rm NL}}

\def\d{{\rm d}}

\def\b{{\vec b}}

\def\T{{\cal T}}

\def\k{{\vec{\scaleto{k}{7pt}}}}

\def\p{{\vec p}}

\def\x{{\vec x}}

\def\y{{\vec y}}

\DeclareRobustCommand{\SkipTocEntry}[4]{}

\newcommand{\hs}{\hspace{1pt}}

\definecolor{colorTC}{rgb}{.2,.7,.2}

\definecolor{amethyst}{rgb}{0.6, 0.4, 0.8}

\definecolor{acolor}{rgb}{0.4, 0.2, 0.4}

\definecolor{blue3}{RGB}{31, 119, 180}
\definecolor{red3}{RGB}{	214, 39, 40}
\definecolor{orange3}{RGB}{255, 127, 14}
\definecolor{green3}{RGB}{44, 160, 44}

\begin{document}

\begin{titlepage}
\setcounter{page}{1} \baselineskip=15.5pt
\thispagestyle{empty}
$\quad$
\vskip 70 pt

\begin{center}
{\fontsize{20}{18} \bf Positivity from Cosmological Correlators}
\end{center}

\vskip 20pt
\begin{center}
\noindent
{\fontsize{12}{18}\selectfont Daniel Green$^1$, Yiwen Huang$^1$, Chia-Hsien Shen$^{1,2,3}$ and Daniel Baumann$^{2,3,4}$}
\end{center}

\begin{center}
\vskip 4pt
\textit{$^1${\small Department of Physics, University of California at San Diego,  La Jolla, CA 92093, USA}}

  \vskip8pt
\textit{$^2$ Department of Physics, National Taiwan University, Taipei 10617, Taiwan}

 \vskip8pt
\textit{$^3$ Leung Center for Cosmology and Astroparticle Physics, Taipei 10617, Taiwan}

  \vskip8pt
\textit{$^4$ Institute of Physics, University of Amsterdam, Amsterdam, 1098 XH, The Netherlands}
\end{center}

%=========================================
\vspace{0.4cm}
 \begin{center}{\bf Abstract}
 \end{center}

\noindent
Effective field theories in flat space and in anti-de Sitter space are constrained by causality and unitarity, often in the form of positivity bounds. Similar bounds have been harder to demonstrate in cosmological backgrounds, where the roles of unitarity and causality are more obscure. Fortunately, the expansion of the universe ensures that late-time cosmological correlators are effectively classical and the role of unitarity is played by classical statistical inequalities. For multi-field inflation, the resulting positivity constraints have long been known in terms of the Suyama-Yamaguchi inequality. In this paper, we demonstrate that similar statistical bounds imply nontrivial constraints for massive fields in the early universe. We show that any real anomalous dimensions for principal series fields in de Sitter space
must be positive.
We also derive a limit on the amplitude of  oscillatory signals from inflation, including those arising in cosmological collider physics. Finally, we demonstrate that these constraints manifest themselves directly in the two-point statistics of matter and galaxies that will be measured in upcoming surveys.

\end{titlepage}
\setcounter{page}{2}

\restoregeometry

\begin{spacing}{1.2}
\newpage
\setcounter{tocdepth}{2}
\tableofcontents
\end{spacing}

\setstretch{1.1}
\newpage

\section{Introduction}

Observables in cosmological spacetimes are far less understood than their counterparts in asymptotically flat or anti-de Sitter spaces. A central challenge in cosmology is the lack of a non-dynamical boundary on which to anchor nonperturbative observables~\cite{Witten:2001kn,Bousso:2004tv}. Furthermore, even on a fixed background, cosmological correlators of quantum fields lack an explicit time variable in which causality and unitarity are usually defined. As a result, it is an open challenge to define the nonperturbative characteristics of well-behaved cosmic observables~\cite{Flauger:2022hie}.

\vskip 4pt
For quantum field theory on a pure de Sitter (dS) background, the isometries act on fields as if they were operators in a Euclidean conformal field theory (CFT) on the future boundary of the spacetime~\cite{Spradlin:2001pw}. Late-time correlators are therefore subject to a familiar set of Ward identities that strongly constrain their functional form. Yet, for theories containing sufficiently massive (principal series) fields, the scaling dimensions of operators are necessarily complex and thus will not obey that same constraints as in a conventional unitary CFT~\cite{Strominger:2001gp}. Defining the allowed space of operators is not only conceptually important, but the dimensions and OPE coefficients of the operators also appear directly in the observable properties of the primordial density fluctuations~\cite{Antoniadis:2011ib,Creminelli:2011mw,Kehagias:2012pd,Bzowski:2012ih,Schalm:2012pi,Mata:2012bx,Kundu:2014gxa,Ghosh:2014kba, Arkani-Hamed:2015bza, Arkani-Hamed:2018kmz}.

\vskip 4pt
A common strategy for constraining the form of physical theories in cosmology is to study their dynamics on small scales where cosmological backgrounds are well-described by their flat-space limit. Microcausality and scattering experiments in this limit are well defined and constrain the microscopic Lagrangian~\cite{Pham:1985cr,Adams:2006sv}. %Equipped with this Lagrangian, t
The implications for inflationary observables can then be derived by direct computations~\cite{Baumann:2011su,Baumann:2015nta,Grall:2020tqc,Green:2022slj,Freytsis:2022aho,deRham:2022hpx}. This strategy has been taken further in the ``cosmological boostrap"~\cite{Baumann:2022jpr}, where the cosmological correlators can be determined from the scattering amplitudes and analyticity.  

\vskip 4pt
The focus on scattering experiments, however, does have its limitations in at least two regards. First, it doesn't constrain the cosmological dynamics directly. Instead, we have to extrapolate the cosmological consequences from the Lagrangian or amplitudes. This limits our ability to infer the outcome of perturbative calculations in cosmology, which have been the source of much confusion in their own right~\cite{Akhmedov:2019cfd,Green:2022ovz}. Second, inflationary backgrounds break  Lorentz boosts and/or the dS isometries~\cite{Cheung:2007st,Baumann:2011su}. We therefore lose the symmetries that play a crucial role in most AdS and flat-space examples (although there has been some recent progress without these assumptions~\cite{Baumann:2019ghk,Pajer:2020wnj,Pajer:2020wxk,Grall:2021xxm,Creminelli:2022onn}). Naturally, we would like to know if there are alternative strategies for understanding the constraints from causality and unitarity directly on cosmological observables.

\vskip 4pt
One general feature of cosmological correlators is that they are described by classical statistics~\cite{Grishchuk:1990bj,Salopek:1990jq,Maldacena:2015bha,Martin:2015qta,Green:2020whw}. This is intuitively clear as the measurement of macroscopic  objects like galaxies should not depend on quantum mechanics. At a technical level, it is also a consequence of the freeze-out of any light field~\cite{dePutter:2019xxv,Cohen:2020php}, as quantum effects are proportional to the decaying mode. Given that classical probabilities are manifestly positive, these observations imply some basic statistical constraints such as the Cauchy-Schwartz inequality and positivity of the Fisher information matrix. 

\vskip 4pt
Statistical constraints of this kind have been derived for the amplitudes of the three- and four-point functions in the squeezed and collapsed limits, in the form of the Suyama-Yamaguchi (SY) inequality~\cite{Suyama:2007bg}. The bounds require that the trispectrum amplitude exceeds the square of the bispectrum amplitude, which arises at tree level from the form of exchange diagrams involving light fields~\cite{Chen:2009zp,Baumann:2011nk,Assassi:2012zq}. Yet, it remains unclear to what extent these inequalities actually carry information about inflationary models\footnote{While the bounds hold in all examples, they may appear to be violated in uncontrolled limits of specific models~\cite{Suyama:2008nt} or in individual loop diagrams~\cite{Sugiyama:2011jt}.}---after all, the bounds are true of any random process regardless of its dynamical origin.

\vskip 4pt
 Not all inflationary four-point functions are manifestly positive. Oscillatory behavior is known to occur in many situations of interest including the exchange of massive fields~\cite{Arkani-Hamed:2015bza}, rapid mixing between fields~\cite{McAneny:2019epy}, chemical potentials~\cite{Bodas:2020yho}, and time-dependent interactions~\cite{Leblond:2010yq}. Statistical bounds typically require four-point functions to be positive, at least in specific limits, which is not obviously the case in these examples. We therefore expect that these models are subject to more subtle constraints to avoid obvious pathologies.
 
\vskip 4pt
In this paper, we will explore information-theoretical constraints on models generating oscillatory bispectra and trispectra. While the approach is a generalization of the proof of the SY inequality~\cite{Smith:2011if}, the implications for the early universe are remarkably different. For massive (principal series) fields in de Sitter space, positivity of the Fisher matrix forbids negative (real) anomalous dimensions. There are only a few direct calculations of these anomalous dimensions~\cite{Marolf:2010zp,Marolf:2010nz} and the constraint on the sign is hardly apparent from Feynman diagrams. Importantly, the Fisher information is positive without any assumptions about symmetries and  the same bounds therefore apply to models of inflation without the conformal-like symmetries of de Sitter space. There one finds that any enhanced oscillatory signal must come together with a larger non-oscillatory contribution that breaks conformal invariance. 

\vskip 4pt
The results derived in this paper arise directly at the level of cosmological observables, including the galaxy power spectra. We show how most of our bounds can be derived within the consistency of the matter and galaxy two-point statistics. Additionally, consistency of the edge cases implies an interesting upper bound on the number density of galaxies. 

\paragraph{Outline}
This paper is organized as follows: In Section~\ref{sec:posd}, we derive positivity bounds on the soft limits of inflationary correlators. In Section~\ref{sec:early_U}, we apply these bounds to fields in de Sitter space and inflation. In Section~\ref{sec:gal}, we show how the same bounds arise in the statistics of galaxies. We conclude in Section~\ref{sec:concl}. Two appendices describe the connection between our classical bounds and quantum-mechanical correlators.

%====================================================
\section{Positivity in Cosmological Correlators}
\label{sec:posd}
%====================================================

Throughout this paper, we will examine the space of allowed cosmological correlators in the limit where they follow from classical statistics. Specifically, bounds on the early universe are derived from constraints on observable adiabatic fluctuations, $\zeta(\k\hs)$, in the superhorizon limit, $k \ll aH$. In this regime, their statistics are necessarily classical and time-independent. As a result, $\zeta$ and any derived quantity\footnote{To avoid confusion later, we will use $\P$ for statistical quantities defined in terms of the late-time observable~$\zeta$, while reserving $\O$ for quantum-mechanical operators related to fields during inflation.} (``operator") $\P(\k\hs)$ will obey the {\it Cauchy-Schwartz inequality} 
\beq
\langle \zeta(\k\hs) \zeta(-\k\hs) \rangle'\, \langle \P(\k\hs) \P(-\k\hs) \rangle' \,\geq\, |\langle \zeta(\k\hs) \P(-\k\hs) \rangle' |^2\,,
\label{equ:CS}
\eeq 
where the prime on the expectation value means that a delta-function has been dropped.
Of course, this inequality easily generalizes to any number of operators  $\P_i$. 

\vskip 4pt
At first sight, the positivity constraint derived from the Cauchy-Schwartz inequality might seem trivial. After all, this must hold for any statistical quantity and appears to be independent of the dynamics of the early universe. However, the fact that cosmic observables freeze out and become classical is a nontrivial statement about the evolution of the universe, which is often not manifest in the perturbative expressions for cosmological correlators. For example, the all-orders proof of the conservation of $\zeta$~\cite{Assassi:2012et,Senatore:2012ya} is highly technical within the usual in-in framework~\cite{Weinberg:2005vy,Weinberg:2006ac}. By using these all-orders results, we will find simple constraints on the  physics of the early universe that are surprisingly challenging to see directly (see Appendix~\ref{app:inin} for additional details).

\vskip 4pt
In this section, we will review how classical statistics constrain the form of inflationary correlators. We will re-derive constraints on non-Gaussianity induced by light fields (the SY inequality) and then generalize the technique to more general types of non-Gaussian statistics. We will apply these general bounds to specific models in Section~\ref{sec:early_U}.

\subsection{Suyama-Yamaguchi and its Generalizations}

In the presence of multiple massless fields, the adiabatic scalar fluctuations $\zeta$ are a local nonlinear function of these additional fields~\cite{Lyth:2002my,Zaldarriaga:2003my,Sasaki:2006kq}. The associated non-Gaussianity is of the ``local type"~\cite{Komatsu:2001rj,Babich:2004gb}:
\begin{align}
\langle \zeta(\k_1) \zeta(\k_2) \zeta(\k_3 ) \rangle' &=\frac{6}{5} \fnl^{\rm local} \left( P(k_1) P(k_2) + P(k_1)P(k_3)+P(k_2) P(k_3) \right) , \\
\langle \zeta(\k_1) \zeta(\k_2) \zeta(\k_3 ) \zeta(\k_4) \rangle' &=  \tnl^{\rm local} \left( P(k_1) P(k_3) P(|\k_1+\k_2|) + {\rm perms} \right) ,
\end{align}
where $P(k) \equiv \langle \zeta(\k\hs) \zeta(-\k\hs)\rangle'$ is the power spectrum, $k \equiv |\k\hs|$, and $\langle \ldots \rangle = \langle \ldots \rangle' (2\pi)^3 \delta_D(\sum \k_i) $. 
The trispectrum contains a sum over 12 terms.
The amplitudes of the bispectrum and trispectrum obey the SY inequality, $\tnl^{\rm local} > (\tfrac{6}{5} \fnl^{\rm local} )^2$~\cite{Suyama:2007bg}.

\vskip 4pt
A proof of the SY inequality~\cite{Smith:2011if} (see also~\cite{Assassi:2012zq,Kehagias:2012pd}) follows directly from the Cauchy-Schwartz inequality~\eqref{equ:CS}. First, we define the operator
\beq\label{eq:zsqr}
\P(\k\hs) \equiv \frac{6\pi^2}{p_{\rm max}^3} \int \frac{\d^3 p}{(2\pi)^3} \frac{\zeta(\p\hs) \zeta(\k-\p\hs)}{P(p)} \, ,
\eeq
where the integration runs over $p\in [0,p_{\rm max}]$. 
The correlations between $\zeta$ and $\P$ can then be written as
\begin{align}
\langle \zeta(\k\hs) \P(-\k\hs)\rangle' &= \frac{6\pi^2}{p_{\rm max}^3} \int \frac{\d^3 p}{(2\pi)^3} \frac{\langle \zeta(\k\hs) \zeta(\p\hs) \zeta(-\k-\p\hs)\rangle'}{P(p)}\, , \\[4pt]
\langle \P(\k\hs) \P(-\k\hs)\rangle' &= \frac{36\pi^4}{p_{\rm max}^6} \int \frac{\d^3 p  \,\d^3p'}{(2\pi)^6} \frac{\langle \zeta(\p\hs) \zeta(\k-\p\hs) \zeta(\p^{\hskip 2pt \prime}) \zeta(-\k-\p^{\hskip 2pt \prime})\rangle'}{P(p) P(p')} \,.
\end{align}
We see that the integrands are given by the bispectrum and trispectrum of $\zeta$.
Assuming local non-Gaussianity, and
taking $p_{\rm max} \gg k$, we can perform the integrals to get
\begin{align}
\langle \zeta(\k\hs) \P(-\k\hs)\rangle' &\to\ \frac{12}{5} \fnl^{\rm local} P(k) \, , \\
\langle \P(\k\hs) \P(-\k\hs)\rangle' &\to\ 4 \tnl^{\rm local} P(k) + \frac{12 \pi^2}{p_{\rm max}^3} \, ,
\label{equ:XXX}
\end{align}
where the second term in (\ref{equ:XXX}) arises from the Gaussian four-point function. The Cauchy-Schwartz inequality (\ref{equ:CS}) then implies
\beq
P(k) \left(  4 \tnl^{\rm local} P(k) + \frac{12 \pi^2}{p_{\rm max}^3} \right) \geq \left( \frac{12}{5} \fnl^{\rm local} P(k) \right)^2 \, .
\eeq
Given that $k \ll p_{\rm max}$, we can neglect the Gaussian term on the left-hand side to find the well-known result~\cite{Suyama:2007bg}
\beq
\boxed{\tnl^{\rm local} \geq \left(\frac{6}{5} \fnl^{\rm local} \right)^2 }\ .
\eeq
This tells us that that power spectrum of $\P$ is positive and bounded from below by its cross-correlation with $\zeta$. 

\vskip 4pt
The SY inequality is easily generalized to non-Gaussian correlators arising from the exchange of light fields during inflation, with masses in the range $0 < m^2 < \frac{9}{4}H^2$ (complementary series). These models, often called {\it quasi-single-field inflation} (QSFI)~\cite{Chen:2009zp}, yield complicated shapes of non-Gaussianity that simplify in the soft limits. Specifically, the squeezed limit of the bispectrum and the collapsed limit of the trispectrum are
\begin{align}
\lim_{k_1 \to 0 } \frac{\langle \zeta(\k_1) \zeta(\k_2) \zeta(\k_3 ) \rangle'}{P(k_1) P(k_2) }  &=  \frac{12}{5} \fnl \left(\frac{k_1}{k_2} \right)^{\Delta} \, , \label{equ:QSFI-1}\\
\lim_{|\k_1+\k_2| \to 0 } \frac{\langle \zeta(\k_1) \zeta(\k_2) \zeta(\k_3 ) \zeta(\k_4) \rangle'}{P(k_1) P(k_3)P(|\k_1+\k_2|)} &=  4\hs \tnl  \left( \frac{|\k_1+\k_2|^2}{k_1 k_3}\right)^\Delta  \, , \label{equ:QSFI-2}
\end{align}
where the scaling parameter $\Delta$ is determined by the mass of the light field 
\beq
\Delta \equiv \frac{3}{2} -\sqrt{\frac{9}{4} - \frac{m^2}{H^2}}\,.
\eeq
Using the operator $\P$ defined in~(\ref{eq:zsqr}), the Cauchy-Schwartz inequality (\ref{equ:CS}) then implies
\beq\label{eq:CS}
P(k) \left(    \frac{36\tnl}{(3-\Delta)^2}\left( \frac{k}{p_{\rm max}}\right)^{2\Delta} P(k) +\frac{12 \pi^2}{p_{\rm max}^3} \right) \geq \left( \frac{36\fnl}{5(3-\Delta)} \left(  \frac{k}{p_{\rm max}}\right)^{\Delta} P(k) \right)^2 \ .
\eeq
For $\Delta < 3/2$, we can again take $p_{\rm max} \gg k$ to find 
\beq
\tnl \geq \left(\frac{6}{5} \fnl \right)^2 \ .
\eeq
It is important to note that we cannot extend this bound to $\Delta \geq 3/2$, because we then can no longer neglect the Gaussian term.

\vskip 4pt
For particles with larger masses,  $m^2 > \frac{9}{4}H^2$  (principal series), the scaling dimensions are complex
\beq
\Delta_\pm = \frac{3}{2} \pm i \sqrt{\frac{m^2}{H^2} - \frac{9}{4}} \equiv \frac{3}{2} \pm i \nu \ .
\eeq
The two  
scaling parameters are complex conjugates, as needed to achieve real solutions for the correlations of $\zeta$. However, the resulting contribution to the trispectrum does not come with a fixed sign. For example, ref.~\cite{Arkani-Hamed:2015bza} found the following trispectrum from the exchange of a massive scalar:
\beq
\lim_{|\k_1+\k_2| \to 0 } \frac{\langle \zeta(\k_1) \zeta(\k_2) \zeta(\k_3 ) \zeta(\k_4) \rangle'}{P(k_1)^{3/2} P(k_3)^{3/2}} = 4\hs \tnl   \left( \kappa(\nu) \left( \frac{|\k_1+\k_2|^2}{k_1 k_3}\right)^{i \nu} + {\rm c.c.} \right) ,
\eeq
where $\kappa(\nu)$ is a complex-valued function of $\nu$. While this trispectrum is not positive, it does {\it not} violate the Cauchy-Schwartz inequality because ${\rm Re}(\Delta) \geq 3/2$ and the Gaussian term in (\ref{eq:CS}) is dominant. However, this example nevertheless points to a potential issue when ${\rm Re} (\Delta) < 3/2$ and ${\rm Im}(\Delta) \neq 0$, that we will explore in detail in Section~\ref{sec:early_U}.

%\subsection{Bispectrum-Weighted Operators}
\subsection{Orthogonal Bispectra Decomposition}

Inflationary models produce bispectra of many possible forms. These are typically defined in terms of an amplitude  and a shape function:
\beq
\langle \zeta(\k_1) \zeta(\k_2) \zeta(\k_3 ) \rangle = \fnl B(k_1,k_2,k_3)\, (2\pi)^3 \delta_D(\k_1+\k_2+\k_3) \,,
\eeq
where we assumed invariance under rotations and spatial translations, and fixed the overall normalization as $B(k_*,k_*,k_*) \equiv 1$ at an arbitrary reference scale $k_*$.  Momentum conservation implies $k_3 =|\k_1-\k_2|$, so we will use $B(k_1,k_2,k_3)$ and $B(\k_1,\k_2)$ interchangeably. Often the space of possible bispectra is parameterized in terms of orthogonal basis functions $B_i(k_1,k_2,k_3)$ (e.g.~\cite{Fergusson:2010dm}), with orthogonality defined by the following inner product
\beq
B_i \cdot B_j = \int \frac{\d^3 k_1\, \d^3 k_2\, \d^3 k_3}{(2\pi)^9} \frac{B_i(k_1,k_2,k_3) B_j(k_1,k_2,k_3)}{P(k_1) P(k_2)P(k_3)} (2\pi)^3 \delta_D(\k_1+\k_2+\k_3)\,.
\eeq
This is a physically meaningful definition~\cite{Babich:2004gb}, as the Fisher matrix for the amplitudes $\fnl^{(i)}$ and $\fnl^{(j)}$ is determined by the same inner product,  $F_{ij} \propto B_i \cdot B_j$.

\vskip 4pt
We will make use of the template decomposition by defining a basis of {\it bispectrum-weighted operators}
\beq\label{eq:P_def}
\P_i(\k\hs) \equiv \frac{6\pi^2}{p_{\rm max}^3} \int \frac{\d^3 p}{(2\pi)^3}  \frac{B_i(\p,\k-\p\hs) \zeta(\p\hs) \zeta(\k-\p\hs)}{P(p)P(k) P(|\k-\p\hs|)} \, .
\eeq
The correlators of $\P_i$ and $\zeta$ are then determined by direct substitution. It is important to recall that the two-point functions of $\P_i$ contain both Gaussian and non-Gaussian contributions 
\begin{align}
\langle \P_i(\k\hs) \P_j(-\k\hs)\rangle' 
&= \frac{36\pi^4}{p_{\rm max}^6} \bigg( \int \frac{\d^3 p}{(2\pi)^3}\frac{2 B_i(\p,\k-\p\hs) B_j(-\p,-\k+\p\hs)}{P(k)^2 P(p) P(|\k-\p\hs|)} \label{eq:template_2pt}  \\[4pt]
& + \int \frac{\d^3 p\, \d^3p'}{(2\pi)^6}  \frac{B_i(\p,\k-\p\hs) B_j(\p^{\hskip 2pt \prime},-\k-\p^{\hskip 2pt \prime})  \langle \zeta(\p\hs) \zeta(\k-\p\hs) \zeta(\p^{\hskip 2pt \prime}) \zeta(-\k-\p^{\hskip 2pt \prime}) \rangle_c'}{P(k)^2 P(p) P(p')P(
|\k-\p\hs|) P(|\k+\p^{\hskip 2pt \prime}|)} \bigg) \, , \nonumber
\end{align}
where $\langle \cdots \rangle_c$ denotes the connected correlator. 
%\db{Cut this QSFI example, which already appeared in the previous section? We show scaling and spin in the next example anyway.} For quasi-single-field inflation, we can then use (\ref{equ:QSFI-1}) and (\ref{equ:QSFI-2}) to get \chs{ $\P_\Delta$ not defined; is it using the choice $B_\Delta = P(k)P(k-p)(k/p)^\Delta$?}
%\chs{Factor of 2 in Eq.~(2.22)? It doesn't reduced to local case when $\Delta=0$.}
%\begin{align}
%\langle \zeta(\k\hs) \P_\Delta(-\k\hs)\rangle' &= \frac{6}{5} \frac{3\fnl}{3-2\Delta} \left(\frac{k}{p_{\rm max}} \right)^{2\Delta} P(k) \,,\\
%\langle \P_\Delta(\k\hs) \P_\Delta(-\k\hs)\rangle' &= \frac{6 \pi^2}{p_{\rm max}^3} \frac{3}{3-2\Delta} \left(\frac{k}{p_{\rm max}} \right)^{2\Delta} + 4 \tnl\left(\frac{k}{p_{\rm max}} \right)^{4\Delta} \frac{9}{(3-2\Delta)^2} P(k) \,,
%\end{align}
%where, as before, we have taken the soft limit of the integrand to perform the integral.

\vskip 4pt
At this stage, it may seem that there is no particular advantage to use the bispectrum-weighted operator. The utility is that it allows us to decompose the bispectrum and trispectrum into a basis of functions, each of which must be positive, much like the K\"all\'en--Lehmann spectral representation. Specifically, with $N$ templates, we have an $(N+1)\times(N+1)$ matrix,
\beq
\mathbb{F} \,\equiv\, \left(\begin{array}{cccc}
\big\langle\zeta(\k\hs) \zeta(-\k\hs)\big\rangle^{\prime} & \big\langle\zeta(\k\hs) \P_1(-\k\hs)\big\rangle^{\prime} &\ldots & \big\langle\zeta(\k\hs) \P_N(-\k\hs)\big\rangle^{\prime} \\[4pt]
\big\langle\P_1(\k\hs) \zeta(-\k\hs)\big\rangle^{\prime} & \big\langle\P_1(\k\hs) \P_1(-\k\hs)\big\rangle^{\prime} &\ldots & \big\langle\P_1(\k\hs) \P_N(-\k\hs)\big\rangle^{\prime} \\[4pt]
\ldots & \ldots & \ldots & \ldots \\[4pt]
\big\langle \P_N(\k\hs)\zeta(-\k\hs)\big\rangle^{\prime} & \big\langle \P_N(\k\hs)  \P_1(-\k\hs) \big\rangle^{\prime} & \ldots & \big\langle \P_N(\k\hs)  \P_N(-\k\hs)\big\rangle^{\prime} 
\end{array}\right)  ,
\eeq
which is positive definite. This construction is qualitatively similar to bounds on EFTs~\cite{Arkani-Hamed:2020blm,Bellazzini:2020cot} and CFTs~\cite{Poland:2022qrs}. We denoted this matrix by $\mathbb{F}$ because it plays the role of the Fisher information matrix for the amplitudes of these operators in the underlying map, at least in the limit where cosmic variance is the dominant source of noise~\cite{Baumann:2021ykm}.

\vskip 4pt
As a concrete illustration, suppose we want to isolate the contributions to the soft limits of cosmological correlators with a specific scaling behavior (mass) and angular dependence (spin). Expanding around the squeezed limit of the bispectrum and the collapsed limit of the trispectrum, we have 
\begin{align}
\lim_{k_1 \to 0}\frac{\langle \zeta(\k_1)\zeta(\k_2) \zeta(\k_3)\rangle'}{P(k_1) P(k_2)} &= \fnl  \sum_{\ell} c_{\ell} \left(\frac{ k_1}{k_2}\right)^{\Delta_\ell} P_\ell(\hat k_1\cdot \hat k_2) \,, \\[4pt]
\lim_{k_I \to 0} \frac{\langle \zeta(\k_1)\zeta(\k_2) \zeta(\k_3) \zeta(\k_4) \rangle_c'}{P(k_1) P(k_3) P(k_I)}  &= \tnl  \sum_{\ell,\ell'}d_{\ell,\ell'}  \frac{P_\ell(\hat k_1 \cdot \hat k_I)P_{\ell'}(\hat k_3 \cdot \hat k_I)}{k_1^{\Delta_\ell} k_3^{\Delta'_{\ell'}} k_I^{-\Delta_\ell-\Delta_{\ell'}'}}  + \cdots \, ,
\end{align}
where $P_\ell(\cos \theta)$ is the $\ell$-th Legendre polynomial\footnote{For the exchange of particles with spin, there will be additional contributions that depend also on $\k_1 \cdot \k_3$ and the associate Legendre polynomials. The generalization to this case is straightforward and is discussed in Section~\ref{sec:dS_anom}.} and $k_I \equiv |\k_1+\k_2|$. 
For simplicity, we have assumed that a single $\Delta_\ell$ dominates for each $\ell$. 

\vskip 4pt
It is useful to introduce a basis of bispectrum templates
\beq\label{eq:Bdl_def}
B_{\ell}(\p,\k-\p\hs) = \left(\frac{k}{p}\right)^{\Delta_\ell} P(p) P(k) P_\ell(\hat k\cdot \hat p ) \, ,
\eeq
to define the operators $\P_{\ell}(\k\hs)$ using~(\ref{eq:P_def}). Since the integrals will be dominated by $p \gg k$, we are only sensitive to the bispectrum template in the soft limit. It is then straightforward to calculate the elements of the matrix ${\mathbb F}$ in this basis
\begin{align}
\langle \zeta(\k\hs) \hs \P_{\ell}(-\k\hs) \rangle' &=\frac{\fnl}{2\ell+1} P(k) \frac{3 c_{\ell}}{3-2\Delta_\ell} \left(\frac{k}{p_{\rm max}} \right)^{2\Delta_\ell} \, , \\[6pt]
\begin{split}\label{eq:ellformula}
\langle \P_{\ell}(\k\hs) \hs \P_{\ell'}(-\k\hs)\rangle' &=   \frac{6 \pi^2}{p_{\rm max}^3} \frac{3}{3-2\Delta_\ell} \left(\frac{k}{p_{\rm max}} \right)^{2\Delta_\ell} \frac{\delta_{\ell,\ell'}}{2\ell+1} \\
&\ \ \ + \frac{\tnl P(k)}{(2\ell+1)(2\ell'+1)}  \left(\frac{k}{p_{\rm max}} \right)^{2\Delta_\ell+2\Delta_{\ell'}'}  \frac{9 d_{\ell,\ell'}}{(3-2\Delta_\ell)(3-2\Delta'_{\ell'})}  \, .
\end{split}
\end{align}
For $\Delta_\ell<3/2$, we can drop the Gaussian term and conclude that $d_{\ell,\ell'}$ is a positive-definite matrix. Isolating specific terms in the Fisher matrix, one also finds that
\beq
\begin{aligned}\label{eq:pos_spin}
d_{\ell,\ell }\, \tnl  &\geq c_{\ell}^2 \fnl^2 \,, \\[4pt]
d_{\ell,\ell } d_{\ell',\ell'} &\geq \left(d_{\ell,\ell'}   \right)^2 \, .
\end{aligned}
\eeq
In this sense, this decomposition of the trispecturm can be diagonized in terms of a matrix will all positive eigenvalues. This can be generalized to the case where $\Delta$ is complex by combining the $\Delta$ and $\Delta^*$ contributions to $c_\ell$ and $d_{\ell,\ell'}$. We will see the utility of this construction in the next section where it is used to isolate contributions to the trispectrum from particles with different masses and spins. 

\vskip 4pt
Naturally, one might hope that the expansion of these correlators using the bispectrum-weighted operators could be used as the starting point for a nonperturbative bootstrap in analogy with CFTs~\cite{Poland:2022qrs}. However, this strategy is more limited than for unitary CFTs, due to the presence of the Gaussian terms which are manifestly positive. Nevertheless, it will be interesting to explore the role of all the minors in the Fisher matrix in this and other applications.

\newpage
\section{Applications to the Early Universe}\label{sec:early_U}

The purpose of this section is to demonstrate that the bounds on the positivity of the Fisher information matrix place nontrivial constraints on the dynamics of the early universe. This is somewhat surprising, as one might expect such bounds to be trivial, in the sense that they must be true for any statistical quantity. However, the late-time observables can encode complicated dynamics of the early universe whose amplitudes and signs are not manifest within the usual perturbative formalism. Our focus in this section will be on processes in the early universe that produce trispectra with oscillating signs. 

\vskip 4pt
The key results in this section are as follows: Scale invariance allows us to classify the long-wavelength behavior of a field in de Sitter or inflation in terms of operators $\O$ with scaling dimensions $\Delta$. In Section~\ref{sec:OPE}, we will show how these operators contribute to the soft limits of cosmological correlators, with oscillatory features occurring when $\Delta$ is complex. Normalizing the two-point functions of such operators as
\beq
\begin{aligned}
\langle \O_{\Delta}(\k\hs) \O_{\Delta}(-\k\hs) \rangle' &\equiv k^{2 \Delta-3} \,, \\[4pt] 
\langle \O_{\Delta}(\k\hs) \O_{\Delta^*}(-\k\hs) \rangle' &\equiv \xi\hs k^{\Delta+\Delta^*-3} \, , 
\end{aligned}
\eeq
we then find that the bounds of Section~\ref{sec:posd} are only satisfied if
\beq
\boxed{\xi \geq 1} \quad {\rm or} \quad \boxed{{\rm Re}(\Delta) \geq \frac{3}{2}} \, .
\eeq
In Section~\ref{sec:dS_anom}, we will show that the de Sitter isometries imply that $\xi=0$ and therefore ${\rm Re}(\Delta) \geq 3/2$ (i.e.~principal series operators in de Sitter must have positive real anomalous dimensions). In Section~\ref{sec:cc}, we will explain how ${\rm Re}(\Delta) < 3/2$ and $\xi \geq 1$ can arise in inflationary models with broken boost symmetry.

\subsection{Soft Limits and OPEs}\label{sec:OPE}

The SY inequality and its generalizations are designed to isolate the squeezed limit of the bispectrum and the collapsed limit of the trispectrum. Assuming that the fluctuations are produced from freeze-out at horizon crossing, these limits capture the influence of  superhorizon (long-wavelength) modes on the statistics of the short modes as they freeze out. 
%In a quasi-de Sitter background, we can organize the theory according to the scaling behavior in $k/(aH)$. In this sense, we can think of 
We will describe this effect in terms of an operator product expansion~(OPE). 

\vskip 4pt
In the limit where $\zeta$ is scale-invariant, the dimension of $\zeta$ is zero and the OPE takes the form~\cite{Green:2020ebl},
\beq
\zeta(\p\hs) \zeta(\k-\p\hs) = P(p)\left[ 1 +\sum_\Delta c_\Delta\, p^{-\Delta} \O_\Delta(\k\hs) + \cdots \right]  ,
\eeq
where $\O_\Delta(\k\hs)$ is a scalar operator of dimension $\Delta$ and the ellipses include operators with spin. We will be particularly interested in the cases where $\Delta$ and $c_\Delta$ are complex. In terms of this expansion, the soft limits of the bispectrum and trispectrum are~\cite{Mirbabayi:2015hva}
\begin{align}
\lim_{k \to 0 } \frac{\langle \zeta(\k\hs) \zeta(\p\hs) \zeta(-\k -\p\hs) \rangle'}{ P(p) }  &=\sum_\Delta c_\Delta \,p^{-\Delta} \langle \zeta(\k\hs) \O_\Delta(-\k\hs) \rangle' \,,\\[4pt]
\lim_{k \to 0} \frac{\langle \zeta(\p\hs) \zeta(\k-\p\hs) \zeta(\p^{\hskip 2pt \prime}) \zeta(-\k-\p^{\hskip 2pt \prime}) \rangle'}{ P(p)P(p')} &=\sum_{\Delta, \Delta'} c_\Delta c_\Delta' \,p^{-\Delta} p'{}^{-\Delta'} \langle \O_\Delta(\k\hs) \O_{\Delta'} (-\k\hs) \rangle' \,.\label{eq:tri_OPE}
\end{align}
If this were a unitary CFT in flat space, then $\Delta$ and $c_\Delta$ would be real, and positivity of~(\ref{eq:tri_OPE}) would follow from the definition of the OPE. In contrast, when $\Delta$ is complex, this expression is not manifestly positive. However, this does not necessarily violate unitarity, since there is no state-operator correspondence in de Sitter~\cite{DiPietro:2021sjt}, despite the similarity to conventional CFTs in terms of the symmetry algebra. See Appendix~\ref{app:two_point} for further discussion.

\vskip 4pt
In single-field inflation, we can choose a gauge where the only degrees of freedom are associated with the metric. As a result, the operators $\O_\Delta$ must then be related to curvatures of the metric and therefore have $\Delta \geq 2$. The resulting constraints on the structure of cosmological correlators are known as the ``single-field consistency condition" and have been well-explored from numerous perspectives~\cite{Maldacena:2002vr,Creminelli:2004yq,Cheung:2007sv,Creminelli:2012ed,Hinterbichler:2012nm,Schalm:2012pi,Assassi:2012zq,Martin:2012pe,Flauger:2013hra}. We will instead be interested in operators with ${\rm Re}(\Delta) \leq 3/2$, which violate the single-field consistency condition.

\subsection{Anomalous Dimensions in de Sitter}\label{sec:dS_anom}

Our first application is the case of de Sitter-invariant fields that are weakly coupled to the
adiabatic perturbation $\zeta$ through an interaction like $\lambda \dot \zeta^2 \sigma$, where $\sigma$ is a local scalar operator in the full theory, such that in the superhorizon limit $\sigma \supset \O_\Delta$. Importantly, since $\zeta$ is real, $\lambda \sigma$ is also real, even though $\O_\Delta$ may be complex. Nevertheless, when $\lambda=0$ and $\dot H / H^2 \to 0$, the correlators of $\sigma$, and hence $\O_\Delta$, are subject to the Ward identities imposed by the dS isometries.

\vskip 4pt
The operators $\O_\Delta$ transform under the dS isometries as primary operators of dimension~$\Delta$:  %in the superhorizon limit. These Ward identities take the form of scale and special conformal transformations, namely
\begin{align}
\O_\Delta(\x) &\to \big(1+ \x \cdot \vec \partial+\Delta\big) \O_\Delta(\x)\,, \\
\O_\Delta(\x) &\to \left[1-2 \Delta \, \x\cdot \b+x^2 \, \b \cdot \vec \partial-2 (\b \cdot \x) \hs \x \cdot \vec{\partial}\hs \right] \O_\Delta(\x) \, ,
\end{align}
where $\vec b$ is an arbitrary real vector. The operators $\O_\Delta$ do not correspond to individual fundamental fields, but do form a basis of operators in the long-wavelength effective description~\cite{Cohen:2020php}.  Applying the Ward identities to the two-point function one finds
\beq\label{eq:cft_2pt}
\langle \O_\Delta(\x) \hs \O_{\Delta'}(\y\hs) \rangle = C_\Delta \hs \delta_{\Delta, \Delta'} |\x-\y\hs|^{-\Delta} + D_\Delta \hs \delta_{\Delta+\Delta', 3} \hs\delta_D(\x-\y\hs) \,,
\eeq
where the second term is a contact term that is allowed when $\Delta +\Delta' =3$.
In the free theory,
 principal series fields have complex $\Delta$, with ${\rm Re}(\Delta) =3/2$, such that $\Delta + \Delta^*=3$ and we can have the contact term~for $\langle \O_\Delta \O_{\Delta^*}\rangle$. Interactions, however, can lead to ${\rm Re} (\Delta) \neq 3/2$ and~(\ref{eq:cft_2pt}) becomes
\beq
\boxed{\langle \O_\Delta(\x) \hs \O_{\Delta^*}(\y\hs) \rangle  = 0}\ ,
\qquad {\rm Re}(\Delta)\neq\frac{3}{2} \, .
\eeq
This result will play a central role in our positivity bounds. It also implies that there are no unitary states associated to operators with ${\rm Re}(\Delta)\neq 3/2$~\cite{Sun:2021thf,Penedones:2023uqc}. It is nonetheless easy to find examples with ${\rm Re}(\Delta) > 3/2$, such as composite operators arising from $\phi^2(\x)$ (see Appendix~\ref{app:two_point}). This is not a contradiction, as there is no state-operator correspondence in de Sitter.

\paragraph{Massive fields}
We will first focus on the case of scalar operators whose dimensions in a free theory would be $\Delta_\pm = \frac{3}{2} \pm i \nu$, with $\nu >0$. When we introduce interactions, the fields can acquire both real and imaginary anomalous dimensions~\cite{Marolf:2010zp,Marolf:2010nz,DiPietro:2021sjt,Mirbabayi:2022gnl,Loparco:2023rug}. The imaginary part is equivalent to a shift in the mass (since it simply shifts the value of $\nu$). We will therefore isolate the real anomalous dimension and define the dimensions of the operators, $\O_\pm \equiv \O_{\Delta_{\pm}}$, as 
\beq
\Delta_\pm = \frac{3}{2} + \gamma \pm i \nu\,,
\eeq
where $\gamma \neq 0$ is real. These dimensions obey $\Delta_+ =\Delta_-^*$ to ensure that the correlators of $\zeta$ are real. Suppose that this operator is coupled to $\zeta$, such that the trispectrum in the collapsed limit takes the form 
\begin{align}
\lim_{k \to 0} \frac{\langle \zeta(\p\hs) \zeta(\k-\p\hs) \zeta(\p^{\hskip 2pt \prime} ) \zeta(-\k-\p^{\hskip 2pt \prime}) \rangle_c'}{P(p)P(p')} = \frac{k^{2\gamma}}{ (p p')^{{3/2}+\gamma}}  \left[ c_+^2 \left(\frac{k^2}{p p'} \right)^{i \nu} + c_-^2 \left(\frac{k^2}{p p'} \right)^{-i \nu} \right] , \label{eq:PS_t}
\end{align}
where we have defined the coupling to $\O_\pm$ in terms of a (quasi)-OPE
\beq
\lim_{k \to 0} \zeta(\p\hs) \zeta(\k-\p\hs) \supset P(p)\Big(c_+\, p^{-\Delta_+}\O_+(\k\hs) + c_-\, p^{-\Delta_-}\O_-(\k\hs) \Big)\,,
\eeq
with the two-point functions normalized as $\langle \O_{\pm}(\k\hs) \O_{\pm}(-\k\hs) \rangle' = k^{2 \Delta_{\pm}-3}$. This normalization is somewhat unconventional, but will greatly simplify our calculations. It is important that the field, in the absence of the coupling to $\zeta$, is dS-invariant, such that $\langle \O_{\Delta_+} \O_{\Delta_-} \rangle = 0$. 

\vskip 4pt
Using~(\ref{eq:zsqr}) to define $\P$, we require that 
\beq
\langle \P(\k\hs)\P(-\k\hs)\rangle' = \frac{36\pi^4}{p_{\rm max}^6} \int \frac{\d^3 p \,\d^3p'}{(2\pi)^6} \frac{\langle \zeta(\p\hs) \zeta(\k-\p\hs) \zeta(\p^{\hskip 2pt \prime}) \zeta(-\k-\p^{\hskip 2pt \prime})\rangle'}{P(p) P(p')} \geq 0\,.
\eeq
Inserting~(\ref{eq:PS_t}) and adding the Gaussian contribution, we get a generalization of~(\ref{eq:CS}): 
\beq\label{eq:CS_PS}
P(k) \left(   \frac{2}{p_{\rm max}^3}\, {\rm Re}\left[\frac{9c_+^2}{(\frac{3}{2} -\gamma - i\nu )^2}\left( \frac{k}{p_{\rm max}}\right)^{2\gamma+ i 2\nu}\right]  +\frac{12 \pi^2}{p_{\rm max}^3} \right) \geq 0 \ .
\eeq
Crucially, because $\nu \neq 0$,
the first term in this expression is sinusoidal and does {\it not} have a definite sign as a function of $k/p_{\rm max}$. Hence, we must have $\gamma \geq 0$, so that this term never dominates the expression. Specifically, taking $\gamma <0$ and $k \to 0$ violates positivity of the power spectrum and therefore only $\gamma \geq 0$ is consistent: 
\beq
\langle \P(\k\hs) \P(-\k\hs)\rangle' \ \xrightarrow[\ p_{\rm max}\gg k\ ]{\gamma < 0} \left(\frac{k}{p_{\rm max}} \right)^{2\gamma} \cos\big[2\nu \log(k/p_{\rm max})\big] \ngeq 0 \quad \implies \quad \boxed{\gamma \geq 0}\ .
\eeq
In other words, unitarity (positive probabilities) requires that any real anomalous dimensions for principal series fields are non-negative.

\vskip 4pt
Positivity of anomalous dimensions has been found in examples by direct computation~\cite{Marolf:2010zp,Marolf:2010nz} and argued for more generally based on the K\"all\'en-Lehmann representation~\cite{Penedones:2023uqc,DiPietro:2021sjt,Mirbabayi:2022gnl,Loparco:2023rug}. Yet, these results depend on technical details about the nature of quantum physics in de Sitter space. Here, we see that the same result holds beyond de Sitter. 
%and is a consequence of the vanishing of $\langle \O_{\Delta_+} \O_{\Delta_-} \rangle = 0$, as required by symmetry in~dS. Specifically, scale invariance would allow a term in~(\ref{eq:PS_t}) of the form $k^{2\gamma} p^{i\nu} p'{}^{-i \nu}$, but violates the special conformal transformations.

\vskip 4pt
\paragraph{Adding light fields} One mechanism for generating anomalous dimensions is through loops of light fields in the complementary series ($0<m^2 < \frac{9}{4}H^2$ ) that interact with the principal series fields~\cite{Marolf:2010zp,Marolf:2010nz}. However, the presence of these additional fields can introduce new contributions to the trispectrum that can potentially relax the bounds on anomalous dimensions. Specifically, combining the results of~(\ref{eq:CS_PS}) and~(\ref{eq:CS}), we require that
\beq \label{eq:osc_tau1}
 \frac{2}{ p_{\rm max}^3}\, {\rm Re}\left[ \frac{9c_+^2}{(\frac{3}{2} -\gamma - i\nu )^2}\left( \frac{k}{p_{\rm max}}\right)^{2\gamma+ i 2\nu}\right] + \frac{36\tnl}{(3-\Delta)^2}\left( \frac{k}{p_{\rm max}}\right)^{2\Delta} P(k) \geq 0\,,
\eeq
where $\Delta$ is the real dimension of the complementary series field. If $\Delta \leq 3/2+\gamma$, positivity can be satisfied as long as $\tnl$ is sufficiently large. Naturally, one might think that if $\gamma < 0$ is always tied to a coupling to complementary series field, this would not be a stringent constraint. However, we will now show that the implications of positivity are more subtle. 

\vskip 4pt
The functional form of the trispectrum contributions in~(\ref{eq:osc_tau1}) for the principal and complementary series fields is quite different. As a result, one might suspect that we can isolate the individual contributions using the right bispectrum-weighted operators. With this in mind, we will consider the following bispectrum template
\beq\label{eq:Bosc}
B(\p,\k-\p\hs) =  P(p)P(k)\, {\rm Re}\left[\left( \frac{p}{k_*} \right)^{\gamma +i \nu}\right] ,
\eeq
for some fixed reference scale $k_*$. 
Using this template to define $\P$ via~\eqref{eq:P_def}, we then have  
\beq
\begin{aligned}
\langle \P(\k\hs) \P(-\k\hs) \rangle' =&  \left(\frac{6\pi^2}{p_{\rm max}^3}\right)^2 \int \frac{\d^3 p\, \d^3 p'}{(2\pi)^6} \,{\rm Re}\left[\left( \frac{p}{k_*} \right)^{\gamma +i \nu}\right] {\rm Re}\left[\left( \frac{p'}{k_*} \right)^{\gamma +i \nu}\right]  \,  \\[4pt]
& \times \frac{P(p)P(p')}{P(|\k-\p\hs|)P(|\k+\p^{\hskip 2pt \prime}|)}\Bigg[ 2G + \frac{k^{2\gamma}}{ (p p')^{3/2+\gamma}}   \bigg( F +   \tau_{\rm NL} P(k) \bigg) \Bigg]  \, , 
\end{aligned}
\eeq
where we assumed $\Delta = 3/2+\gamma$ for the complementary series field and defined
\begin{align}
F &\equiv c_+^2 \left(\frac{k^2}{p p'} \right)^{i \nu} + c_-^2 \left(\frac{k^2}{p p'} \right)^{-i \nu}\,, \\[4pt]
G &\equiv (2\pi)^3 \delta_D(\p+\p^{\hskip 2pt \prime}) \,.
\end{align}
These integrals are straightforward, but messy to evaluate. However, when $\nu \gg 1$, the results simplify and yield
\beq
 \frac{2}{p_{\rm max}^3}\, {\rm Re}\left[ c_+^2\left( \frac{k}{k_*} \right)^{2\gamma+ i 2\nu}+ \frac{18\pi^2}{3+2 \gamma}\left(\frac{p_{\rm max}}{k_*}\right)^{2\gamma}\right] + O\left(\nu^{-1} \tnl \right)\geq 0 \, .
 \label{equ:XXXX}
\eeq
In the high-mass limit, $\nu \gg 1$, we again find that $\gamma <0$ is forbidden by the positivity of the Fisher information matrix, even when we include $\tnl > 0$ to make the trispectrum positive in the sense of~(\ref{eq:osc_tau1}). Specifically, the bispectrum template suppresses this contribution, although it does not project it out entirely.\footnote{It is possible we could isolate the principal series term with a different choice of bispectrum template.} Nevertheless, we see that adding complementary series fields does not offer a trivial resolution to the apparent violation of unitarity associated with $\gamma < 0$, although this may be possible for $\nu = {\cal O}(1)$. This example also highlights the utility of the template decomposition of $\P$ to isolate the different kinds of operators contributing to the trispectrum.

\paragraph{Particles with spin} Every quantum field theory comes with a tower of operators of different spins. These operators are essential for the description of correlators in terms of an OPE and many other applications. In AdS/CFT, these operators are particularly important as their anomalous dimensions and OPE coefficients in the CFT are in one-to-one correspondence to the dynamics in (and of) the bulk spacetime~\cite{Heemskerk:2009pn,Fitzpatrick:2010zm}. Naturally, this suggests that understanding these operators could have similar implications for understanding the dynamics of dS. In addition, particles with spin offer a unique signal of interest for cosmological collider physics~\cite{Arkani-Hamed:2015bza,Lee:2016vti}. It therefore is important to  extend the results of the previous section to operators with spin $s>0$.

\vskip 4pt
In a CFT, the two-point functions of operators with spin $s$ are~\cite{Simmons-Duffin:2016gjk}
\beq\label{eq:2pt_spin}
\left\langle [\O_\Delta]^{\mu_1 \ldots \mu_{s}}(x) [\O_\Delta]_{\nu_1 \ldots \nu_{s}}(0)\right\rangle=C_{s, \Delta} \left(\frac{I^{(\mu_1}{}_{(\nu_1}(x) \cdots I^{\mu_{s})}{ }_{\nu_s)}(x)}{x^{2 \Delta}}-\text { traces }\right) ,
\eeq
where $I^\mu{ }_\nu(x) \equiv \delta_\nu^\mu-2 \hs x^\mu x_\nu/x^2$ and the scaling dimensions (for a free field) are 
\beq
\Delta_{ \pm}=\frac{3}{2} \pm \sqrt{\left(s-\frac{1}{2}\right)^2-\frac{m^2}{H^2}}  \ .
\eeq
When $\Delta_\pm$ is real, the two-point functions for physical operators must be positive, which leads to the unitarity constraint $m^2=0$ or $m^2 \ge s(s-1)H^2$~\cite{Higuchi:1986py,Deser:2001us}. 
For $m^2 > (s-\frac{1}{2})^2 H^2$, we have $\Delta_{\pm} = \frac{3}{2} \pm i \nu$, which allows a delta-function solution to $\langle \O_\Delta \O_{\Delta^*}\rangle$ as in (\ref{eq:cft_2pt}).

\vskip 4pt
The exchange of spinning operators leads to the following
 trispectrum~\cite{Arkani-Hamed:2015bza}: 
\begin{align}
\lim_{k \to 0} \frac{\langle \zeta(\p\hs) \zeta(\k-\p\hs) \zeta(\p^{\hskip 2pt \prime}) \zeta(-\k-\p^{\hskip 2pt \prime}) \rangle_c'}{P(p)P(p')} &= \frac{k^{2\gamma} }{ (p p')^{3/2+\gamma}} \left[ S_+ c_+^2 \left(\frac{k^2}{p p'} \right)^{i \nu} + S_- c_-^2 \left(\frac{k^2}{p p'} \right)^{-i \nu} \right] ,
\end{align}
which is a generalization of \eqref{eq:PS_t} that includes an
angular dependence through the functions $S_\pm$. The latter can be written as
\beq
S_\pm(\hat k,\hat p, \hat p') \equiv \sum_{m=-s}^s e^{i m\left(\psi-\psi^{\prime}\right)} P_s^m(\cos \theta) P_s^m\left(\cos \theta^{\prime}\right) a^{(s,m)}_\pm\,,
\eeq
where $P_s^m(x)$ are associated Legendre polynomials, with $\cos \theta \equiv \hat k \cdot \hat p$, $\cos \theta' \equiv \hat k \cdot \hat p'$, 
and $\psi-\psi'$ being the angle between $\p$ and $\p^{\hskip 2pt \prime}$ in the plane orthogonal to $\k$. In principle, one can find all the coefficients $a^{(s,m)}_\pm$, as a function of $\Delta_\pm$, by Fourier transforming~(\ref{eq:2pt_spin}).  For the tree-level exchange of particles with spin, these solutions are known~\cite{Arkani-Hamed:2015bza}.

\vskip 4pt
Using orthogonality of the Legendre polynomials, we can isolate any given term in the sum over helicities. For example, for $m=0$, we can use~(\ref{eq:ellformula}) and~(\ref{eq:pos_spin}) to find 
\beq
  {\rm Re}\left[ \frac{3\hs a^{(s,0)}_+}{(2s+1)(\frac{3}{2} -\gamma - i\nu ) }\, c_+^2 \left( \frac{k}{p_{\rm max}}\right)^{2\gamma+ i 2\nu}\right] + 3 \pi^2  \geq 0 \, ,
\eeq
which is a generalization of (\ref{equ:XXXX}).
This is now identical to the scalar case: when $\gamma < 0$, we can neglect the constant as $k\to 0$. However, the resulting correlator does not satisfy the bound~(\ref{eq:pos_spin}) and we thus require for consistency that\footnote{Technically, it only forces the $m=0$ component to vanish. However, repeating the analysis for $m \neq 0$, we arrive at the conclusion that $\gamma \geq 0$.} 
\beq
%\hspace{-0.1cm}\langle \P_\ell(\k\hs) \P_\ell(-\k\hs)\rangle^\prime  \xrightarrow[\ p_{\rm max}\gg k\ ]{\gamma < 0} \frac{ | a^{(s,0)}_+ c_+^2|}{2s+1} \left(\frac{k}{p_{\rm max}} \right)^{2\gamma} \cos\big[2\nu \log\left({k/p_{\rm max}}\right)\big] \ngeq  0  
 %\implies 
 \boxed{\gamma \geq 0} \,  . 
 \eeq
In this sense, there is no meaningful distinction between the bounds on heavy scalars and heavy particles with spin.

\subsection{Signals from the Cosmological Collider}\label{sec:cc}

During inflation, the time-dependent background breaks the de Sitter isometries and allows for non-Lorentz-invariant interaction terms~\cite{Cheung:2007sv,Baumann:2011su}. It is then possible to find multi-field models where a field has $\Delta_\pm = r\pm i \nu$, with $r \ll 1$ and $\nu >1 $. Such models would have a large non-Gaussian signal with a unique oscillatory shape. In the previous section, we proved that this is impossible in exact de Sitter ($r \geq 3/2$), but we will now identify precisely how this can be evaded within inflationary models. Specifically, we will show that such a feature is possible only if it is accompanied by a larger non-oscillatory contribution to the trispectrum.

\paragraph{A model with $\gamma <0$}  While we may naturally guess that the bounds from dS are weakened in the context of inflation, it is another matter to find specific examples that saturate the weaker bounds. In particular, given the technical challenges in calculating anomalous dimensions via loops, we would like to have a simple example where $\gamma < 0$ unambiguously. 

\vskip 4pt
A straightforward mechanism for achieving $\gamma <0$ is through the kinetic mixing of two massive scalar fields $\sigma_{1,2}$~\cite{McAneny:2019epy}: 
\beq\label{eg:neg_eg}
{\cal L}_{\sigma} =- \frac{1}{2}  \sum_{i=1}^2\left(\partial_\mu \sigma_i \partial^\mu \sigma_i - m_i^2 \sigma_i^2\right) + \rho (\dot\sigma_1 \sigma_2 -\dot \sigma_2 \sigma_1) \, ,
\eeq
where the parameter $\rho \propto \dot \phi$ breaks Lorentz invariance, with $\phi(t)$ being the time-dependent inflaton. The resulting equations of motion are
\begin{align}
\ddot \sigma_{1} + 3 H \dot \sigma_{1} + \left(\frac{k^2}{a^2}+m_1^2\right)\sigma_{1} + \rho (2 \dot \sigma_{2}+ 3 H \sigma_{2}) &= 0\,, \\
\ddot \sigma_{2} + 3 H \dot \sigma_{2} + \left(\frac{k^2}{a^2}+m_2^2\right)\sigma_{2} - \rho (2 \dot \sigma_{1}+ 3 H \sigma_{1}) &= 0\,.
\end{align}
Our primary interest is understanding the allowed values of $\Delta$ in the limit $k \to 0$. Writing $\vec \sigma = (\sigma_1, \sigma_2)= a(t)^{-\Delta} \vec v$, where $\vec v$ is a constant vector, the equations of motion for $k=0$ and $\dot H \to 0$ become
\beq\label{eq:qsfi_matrix}
\begin{pmatrix}
\Delta(\Delta - 3)H^2 + m_1^2 & H \rho( -2 \Delta+3)  \\[4pt]
H \rho( 2 \Delta-3) & \Delta(\Delta - 3)H^2 + m_2^2 
\end{pmatrix} \, \vec v = 0 \, .
\eeq
Solving this equation is equivalent to finding the values of $\Delta$ for which the matrix has a vanishing determinant.
The resulting quartic equation has the following four solutions:
\beq
\Delta_{\pm} = \frac{3}{2} - \sqrt{\frac{9}{4}+ a_\pm} \quad {\rm and} \quad 3 -\Delta_\pm\,,
\eeq
where
\beq
a_\pm \equiv -\frac{1}{2} \frac{m_1^2+ m_2^2 +4 \rho^2 }{H^2} \pm 2 \sqrt{\frac{1}{2}\frac{\rho^2}{H^2}\left(\frac{m_1^2 + m_2^2 + 2\rho^2}{H^2}  -\frac{9}{2} \right)+\frac{1}{16}\frac{(m_1-m_2)^2}{H^2}} \, .
\eeq
Our goal then is to find solutions of the form $\Delta = r + i \nu$, with $r < 3/2$ and $\nu \neq 0$. Clearly, if $a_\pm$ is real, then we will have either $\nu=0$ or $r=3/2$ for $a_\pm \geq -9/4$ and $a_\pm < -9/4$, respectively. The only regime where $r\neq 0$, but our solutions are complex, is when $a_\pm$ is itself complex. This is only possible when 
\beq
\frac{1}{2}\frac{m_1^2 +m_2^2 + 2\rho^2}{H^2}  + \frac{1}{16}\frac{(m_1-m_2)^2 }{\rho^2}< \frac{9}{4} \, .
\eeq
There are regimes of parameter space where this inequality is satisfied and the solutions are stable. For example, taking $\rho =H$ and $m_1^2=m_2^2=0.1\,H^2$ gives $\Delta_\pm = 0.42 \pm i$. On the other hand, notice that taking $m_i^2\gg \rho^2 \gg H^2$ always implies that ${\rm Re}(\Delta_\pm) = 3/2$, as one would expect from the redshifting of massive particles in an expanding universe~\cite{Arkani-Hamed:2015bza}.

\paragraph{Positivity bounds}
The above example clearly demonstrates that $\gamma < 0$ is allowed if we break the symmetries of de Sitter space. Although we can quantify the conditions for this to occur in the model, we would like to derive a model-independent bound on the amount and type of symmetry breaking needed in any self-consistent example. To do this, we will return to our OPE description.

\vskip 4pt
Let us assume an inflationary model that includes operators $\O_\pm$ whose scaling dimensions (in the superhorizon limit) are $\Delta_\pm = r \pm i\nu$. Coupling these operators to $\zeta$ gives the following trispectrum (in the collapsed limit)
\beq
\begin{aligned}
\lim_{k \to 0} \frac{\langle \zeta(\p\hs) \zeta(\k-\p\hs) \zeta(\p^{\hskip 2pt \prime}) \zeta(-\k-\p^{\hskip 2pt \prime}) \rangle'}{P(p)P(p')} &= \frac{k^{2r - 3}}{ (p p')^{r}}  \left(c_+ c_- \xi \left(\frac{p}{p'} \right)^{i \nu} + c_+^2 \left(\frac{k^2}{p p'} \right)^{i \nu} + {\rm c.c.} \right) \label{eq:cc_bi}  ,
\end{aligned}
\eeq
where we defined
\beq
\langle \O_+(\k\hs) \O_-(-\k\hs) \rangle' = \frac{\xi}{k^3} k^{\Delta_++\Delta_-} = \xi k^{2r -3 }  \ .
\eeq
Unlike in de Sitter, $\xi\neq 0$ is now allowed because the inflaton defines a preferred slicing and therefore breaks to dS isometries.

\vskip 4pt
We will again define the bispectrum-weighted operator $\P$ using $B(\p, \k-\p\hs)$ from~(\ref{eq:Bosc}) and replacing $r = 3/2 + \gamma$. We are only interested in the regime $\gamma < 0$ and we can therefore drop the Gaussian contribution and calculate 
\beq
\begin{aligned}
&\langle \P(\k\hs) \P(-\k\hs) \rangle' = \left(\frac{6\pi^2}{p_{\rm max}^3}\right)^2 \int \frac{\d^3 p\, \d^3 p'}{(2\pi)^6} \, {\rm Re}\left[\left( \frac{p}{k_*} \right)^{\gamma +i \nu}\right] {\rm Re}\left[\left( \frac{p'}{k_*} \right)^{\gamma +i \nu}\right]\,\times \\[4pt]
&\quad \times \frac{P(p)P(p')}{P(|\k-\p\hs|)P(|-\k-\p^{\hskip 2pt \prime}|) } \frac{k^{2\gamma}}{(p p')^{3/2 + \gamma}} \left(c_+ c_-\, \xi \left(\frac{p}{p'} \right)^{i \nu} + c_+^2 \left(\frac{k^2}{p p'} \right)^{i \nu} + {\rm c.c.} \right)  .
\end{aligned}
\eeq
Taking $\nu \gg 1$ and dropping terms that are suppressed by powers of $\nu$, we find 
\beq\label{eq:kappa_bound}
2 \,\frac{|c_+|^2 \xi}{p_{\rm max}^3} \left( \frac{k}{k_*} \right)^{2\gamma} +  2\, {\rm Re}\left[ \frac{c_+^2}{p_{\rm max}^3}\left( \frac{k}{k_*} \right)^{2\gamma+ i 2\nu}\right] \geq 0 \quad \implies \quad \boxed{\xi \geq 1} \ .
\eeq
We can think of this bound as the minimal level of conformal-symmetry breaking that is necessary to arrive at $r < 3/2$ (i.e.~$\gamma <0$).

\paragraph{An example} To satisfy the positivity constraint in~(\ref{eq:kappa_bound}), the amplitudes of the various contributions to the trispectrum must be related. The model in~(\ref{eg:neg_eg}) gives us a concrete demonstration of how such correlations can arise. Concretely, we recall that the scaling solutions to~(\ref{eq:qsfi_matrix}) must match onto the full solutions to the mode function, $f_i(\k, a(t) )$, defined by the $\hat \sigma_i$ operators
\beq 
\hat \sigma_i(\k,a(t)) = \hat a^{(i)}_\k f_i(\k, a(t) )+ {\rm h.c.} \, ,
\eeq
where $\hat a_\k^{(i)}$ are annihilation operators. In the limit $k/(aH) \to 0$, the mode functions must reduce to a combination of these solutions,
\beq
f_i(\k\hs)= \sum_\Delta c_\Delta (\vec v_\Delta)_i \frac{k^\Delta}{a^{\Delta} k^{3/2}} \, .
\eeq
In this way, $f_i(\k\hs)$ contains all the contributions from the scaling operators of the long-wavelength solutions, $\O_\Delta(\k)$. However, $\zeta$ cannot couple individually to $\O_\Delta$; it can only couple locally to $\sigma_i(\x)$. Since $\vec \sigma(\x)$ is a real field, it contributes to the trispectrum via an exchange diagram that is proportional to its power spectrum, 
\beq
\langle \sigma_i(\k\hs) \sigma_i(-\k\hs) \rangle' =  |f_i(k,a)|^2  > 0 \, .
\eeq
 We have also verified that our positivity constraints are obeyed for the bispectra and trispectra calculated in the explicit examples of $\gamma <0$ found in~\cite{McAneny:2019epy}.  In both cases, the correlations needed to satisfy~(\ref{eq:kappa_bound}) occur because the operators $\O_\pm(\k)$ are defined in terms of a single pair of creation and annihilation operators $\hat a^\dagger_\k$ and $\hat a_{-\k}$.

\vskip 4pt
In Appendix~\ref{app:two_point}, we show that the absence of manifest positivity for dS correlators when $\gamma \neq 0$ requires a failure of a reality condition for the operator in the exchange diagram. Even for $\gamma > 0$, the implication is that loop corrections must spoil these conditions. There is no such issue here, as $\gamma \neq 0$ arises as a solution to the quadratic field equations and reality (and therefore positivity) is manifest via canonical quantization.

\section{Positivity in Galaxy Surveys}\label{sec:gal}

In this section, we will show how the positivity bounds on the primordial statistics manifest themselves in the late-time distribution of galaxies. We will see that some of our bounds can be derived directly from the correlations of galaxies and matter. Yet, consistency between the positivity of the primordial and late-time statistics also constrain the physics of galaxy formation.

\subsection{Review of Scale-Dependent Bias}

Nontrivial squeezed and collapsed limits of the primordial statistics introduces long-range correlations in the distribution of galaxies and other collapsed objects~\cite{Dalal:2007cu} (see also~\cite{Schmidt:2010gw,2012JCAP...08..019N,Baumann:2012bc,Pajer:2013ana,Assassi:2015fma,Gleyzes:2016tdh}). The resulting enhancement of the galaxy power spectrum at long wavelengths, termed {\it scale-dependent bias}, is one of the key observational tools being used to search for non-Gaussian signals from inflation~\cite{SPHEREx:2014bgr,DESI:2022lza,Schlegel:2022vrv}.
Following~\cite{Baumann:2012bc}, we will briefly review how the primordial correlations are imprinted in this scale-dependent bias.

%Following~\cite{Baumann:2012bc}, we will first review how deviations from the single-field consistency conditions manifest themselves in the distributions of halos and galaxies.\footnote{See e.g.~\cite{Creminelli:2013mca,Creminelli:2013poa,Pajer:2013ana,dePutter:2015vga} for derivations of the single-field consistency conditions directly at the level of galaxies or other tracers of the large-scale structure.} 

\vskip 4pt
At linear order, the relationship between the primordial density fluctuations $\zeta(\x)$ and the (fractional) matter overdensity $\delta_m(\x)$ is given by
\beq
\delta_m(z,\k\hs) = \frac{k^2 T(k) D(z)}{\Omega_m H^2} \zeta(\k\hs) \equiv \T(k,z) \zeta(\k\hs) \,,
\eeq
where $z$ is the redshift, $T(k)$ is the linear transfer function and $D(z)$ is the growth factor.

\vskip 4pt
For our purposes, it is sufficient to assume that the number of collapsed objects is determined by the local gravitational evolution, and hence the local matter density fluctuations. In other words, the number density of galaxies, $n_g(\x)$, is some general function of local composite operators made from $\delta_m(\x)$ and its derivatives~\cite{2009JCAP...08..020M}. Defining the galaxy overdensity as $\delta_g(\x) \equiv (n_g(\x)-\bar n_g) /\bar n_g$ and expanding it in powers of $\delta_m$, we have the following {\it bias expansion}
\beq
\delta_g(\x) = b_1 \hs \delta_m(\x) + b_2 \hs \delta_m^2(\x)+ \cdots +\epsilon(\x)  \, ,
\eeq
 where the parameters $b_n$ are the bias coefficients and $\epsilon(\x)$ is a ``stochastic bias" that is uncorrelated with $\delta_m(\x)$ (see e.g.~\cite{Desjacques:2016bnm} for review). The stochastic bias is required because  $n_g(\x)$ is a discrete variable and $\delta_m(\x)$ is continuous. 

\vskip 4pt
Higher-order operators in the bias expansion
encode non-Gaussian statistics in the distribution of galaxies.  For example,  the operator
\beq
[\delta^2_m](z,\k\hs) = \int \frac{\d^3 p}{(2\pi)^3} \,\T(p,z) \T(|\k-\p\hs|,z) \zeta(\p\hs)\zeta(\k-\p\hs) 
\eeq
leads to a
 cross correlation of galaxies and matter $P_{gm}(k) \equiv \langle \delta_g \delta_m \rangle'$ that depends on  the primordial bispectrum, while the galaxy power spectrum $P_{g}(k) \equiv \langle \delta_g \delta_g \rangle'$ contains the primordial trispectrum. 
As a result, we get 
\begin{align}
P_{gm}(k) &\,\supset\,  b_\phi \fnl \left(k R_*\right)^{\Delta} \T(k) P(k)\ \, =   b_\phi \fnl \frac{\left(k R_*\right)^{\Delta}}{\T(k)} P_m(k)  \, , \\
P_{g}(k)  &\,\supset\,  b^2_\phi \left(\frac{5}{6} \right)^2 \tnl \left(k R_*\right)^{2\Delta}  P(k) =   \left(\frac{5}{6} \right)^2 \tnl b^2_\phi  \frac{\left(k R_*\right)^{2\Delta}}{\T(k)^2} P_m(k) + P_\epsilon(k) \, ,
\end{align}
where $P_\epsilon(k) = \langle \epsilon(\k\hs) \epsilon(-\k\hs) \rangle'$. The parameter $b_\phi$ is  associated with the regulator $R_*^{-1} \sim p_{\rm max}$ in the integrals over $p$ or $p'$. In practice, the precise value $b_\phi$ depends on $\Delta$, but also on mass, redshift, and the (unknown) physics of galaxy formation. These details are important for precise experimental constraints on $\fnl$ and $\tnl$, but are unimportant for the positivity of the Fisher information. See e.g.~\cite{Baumann:2012bc}, for more details.

\vskip 4pt
For models of inflation with ${\rm Re}(\Delta) < 3/2$ and ${\rm Im}(\Delta) \neq 0$, like those discussed in Section~\ref{sec:cc}, the cross correlation is purely oscillatory~\cite{McAneny:2019epy} 
\beq\label{eq:complexgm}
P_{gm}(k)\, \supset\,  b_{\phi}  \fnl \left(k R_*\right)^{{3/2} +\gamma} \cos\Big( \nu \log k R_* +\varphi \Big)\, \T(k) P(k) \, ,
\eeq
where we used~(\ref{eq:cc_bi}) and defined a phase $\varphi$ which depends ${\rm arg}(c_+)$ and on the details of the biasing model. Meanwhile, the galaxy power spectrum receives both oscillatory and non-oscillatory contributions
\beq\label{eq:complexgg}
P_g(k) \,\supset\,  b_{\phi}^2 \left(\frac{5}{6} \right)^2 \tnl \left(k R_*\right)^{3 +2\gamma}\bigg[ \xi + \cos\Big( 2 \nu \log k R_* +\varphi' \Big) \bigg] P(k)  \, ,
\eeq
where the phase $\varphi'$ is not necessarily the same as $\varphi$. The bound $\xi \geq 1$ ensures positivity of the power spectrum as $k \to 0$. This is the same as our bound from the Fisher information in~(\ref{eq:kappa_bound}). Here, we have allowed for the possibility of $\fnl \neq 0$ through an additional mixing of $\zeta$ and $\sigma$ via an interaction like $\lambda' \dot \zeta \sigma$ (in addition to the existing $\lambda \dot \zeta^2 \sigma$ interaction). See~\cite{McAneny:2019epy}, for a detailed example.

\subsection{Positivity and the Galaxy Power Spectrum}

The statistics of galaxies is a hallmark of the kind classical statistical quantity on which our positivity bounds were derived. It is also precisely the observable that will drive future constraints on non-Gaussianity. We would therefore like to understand in what sense our previous bounds could be derived directly from the statistics of galaxies, rather than through our fictitious operators $\P_i(\k\hs)$.

\vskip 4pt
In the late universe, $\delta_m$ and $\delta_g$ are classical stochastic variables and therefore obey the Cauchy-Schwartz inequality
\beq
P_{m}(k) P_{g}(k) \geq |P_{mg}(k)|^2 \, .
\eeq
Applying this to the conventional scale-dependent bias discussed in the previous subsection, we get
\beq
b_\phi^2 \left[\left(\frac{5}{6} \right)^2 \tnl-\fnl^2\right]  \frac{\left(k R_*\right)^{2\Delta}}{\T(k)^2} P_m(k) + P_\epsilon(k)\geq 0 \, .
\eeq
To isolate the non-Gaussian term, we take the limit $k\to 0$. In this limit, the stochastic term is simply a constant that is well approximated by shot noise, $P_\epsilon(k) \approx {\bar n_g}^{-1}$. In contrast, $P_m(k\to 0) \propto k$ and therefore the matter power spectrum itself is suppressed relative to the shot noise. However, the non-Gaussian term is enhanced as $k\to 0$ since $\T(k) \to k^2$ in that limit. Using 
\beq
\frac{1}{\T(k)^2 }P_m(k) = P(k) \equiv \frac{A_s}{k^3} \ ,
\eeq
we find
\beq
\lim_{k\to 0} \left(b_\phi^2\left[\left(\frac{5}{6} \right)^2 \tnl-\fnl^2\right]  \left(k R_*\right)^{2\Delta} \frac{A_s}{k^3}  + \frac{1}{\bar n_g} \right) \geq 0 \, .
\eeq
As a result, just like in the primordial statistics, we find that positivity of the galaxy correlators implies the SY inequality $\tnl\geq (\tfrac{6}{5}\fnl)^2$, if we assume $\Delta < 3/2$.

\vskip 4pt
Interestingly, for $\Delta= 3/2$, we can still derive a bound  
\beq
b_\phi^2 R_*^3 \left[ \left(\frac{5}{6} \right)^2 \tnl-\fnl^2\right]  A_s \geq - \frac{1}{\bar n_g} \, .
\eeq
At face value, this does not place an interesting constraint on the correlators. However, it must also be true that the correlators of galaxies cannot contain more information than the primordial statistics~\cite{Baumann:2021ykm}. For $\Delta=3/2$, the bound from the initial conditions is 
\beq
 \left[ \left(\frac{5}{6} \right)^2 \tnl-\fnl^2\right]  A_s \geq - \frac{3 \pi^2}{4}\,.
\eeq
Requiring the galaxy-based bound not to exceed this bound from the initial conditions then implies that
\beq\label{eq:ng_bound}
 \boxed{\bar n_g \leq \frac{4}{3\pi^2} \frac{1}{b_\phi^2 R_*^3} }\ .
\eeq
This upper-limit on $\bar n_g$ makes sense intuitively. We do not expect every region of the characteristic scale of a halo, $R_*^3$, to contain a galaxy. This physics expectation roughly translates to $n_g R_*^{-3} < 1$. Nonetheless, the bound~(\ref{eq:ng_bound}) is a strict requirement that includes the parameters $b_\phi$ and $R_*$, which set the amplitude of the signal for any $\tnl$ and $\fnl$.

\vskip 4pt
Finally, we can repeat this analysis for the oscillatory bispectra and trispectra. The two-point statistics, $P_{mg}(k)$ and $P_{g}(k)$, are given by~(\ref{eq:complexgm}) and~(\ref{eq:complexgg}), respectively. Assuming $\gamma < 0$, so that we can neglect the stochastic term, the Cauchy-Schwartz inequality implies 
\beq
\begin{aligned}
 b_{\phi}^2 &\left(\frac{5}{6} \right)^2 \tnl \Big(k R_*\Big)^{3 +2\gamma}  \bigg[ \xi  +\cos\Big( 2 \nu \log k R_* +\varphi' \Big) \bigg] P_m(k) P(k) \\
&\geq\ 
\bigg[  b_{\phi}  \fnl \left(k R_*\right)^{{3/2} +\gamma} \cos\Big( \nu \log k R_* +\varphi \Big)\bigg]^2 P_m(k) P(k) \, . 
\end{aligned}
\eeq
A sufficient condition for satisfying this inequality is
\beq
\tnl \big(\xi \pm 1\big) \geq \left(\frac{6}{5} \fnl \right)^2 \, . 
\eeq
As a purely theoretical statement, this implies that, as $k\to 0$, there will be a non-oscillatory contribution to $P_{gg}(k)$ whose amplitude is larger than the oscillatory signal. In practice, observational constraints on non-Gaussianity in the near-term arise from the regime at finite $k$ where 
\beq
P_{g}(k) \approx \left[ b_1^2 + 2 b_1 b_{\phi}  \frac{\fnl}{ \T(k)} \left(k R_*\right)^{{3/2} +\gamma} \cos\Big( \nu \log k R_* +\varphi \Big) \right] P_m(k) \, .
\eeq
As a result, it is entirely possible that the oscillatory contribution to the scale-dependent bias will be the dominant observational signal in the galaxy power spectrum at small but finite $k$. This simply reflects the fact that the signal-to-noise of the bispectrum is larger than that of the trispectrum when $\tnl = O(\fnl^2)$.

\section{Conclusions}\label{sec:concl}

The nature of quantum field theory and quantum gravity in accelerating cosmologies have long been a source of confusion. Without a clear definition of nonperturbative observables, progress has relied on perturbative calculations. Unfortunately, these calculations are themselves technically challenging and have led to much confusion~\cite{Green:2022ovz}. Straightforward constraints on dynamics in these spacetimes is an important tool in the path towards understanding the quantum nature of our universe. Intuition suggests that on small scales, flat-space consistency conditions will still apply. Yet, deriving strict bounds in the cosmological setting has proven challenging.

\vskip 4pt
Fortunately, long-wavelength fluctuations in these accelerating spacetimes are subject to familiar symmetries. Both de Sitter space and inflationary background possess an $SO(d,1)$ symmetry that is either linearly or non-linearly realized~\cite{Creminelli:2012ed,Hinterbichler:2012nm}. Organizing observables according to these symmetries has driven process in both the cosmological bootstrap~\cite{Baumann:2022jpr} and EFT approaches to cosmic observables~\cite{Green:2022ovz}. 

\vskip 4pt
Yet, despite these symmetries, the behavior of fields in de Sitter space has long been known to evade the usual constraints  of a unitary CFT. This has limited the utility of the dS/CFT correspondence~\cite{Strominger:2001pn,Maldacena:2002vr} as a tool for understanding de Sitter space. Cosmic observables on a fixed time-slice are Euclidean and thus not required to obey the traditional unitarity constraints. Naturally, one is left to ask how the rules of unitarity manifest themselves~\cite{Arkani-Hamed:2018kmz,Melville:2021lst,Goodhew:2021oqg}, if at all~\cite{Cotler:2022weg,Cotler:2023eza}, in these backgrounds. Furthermore, constraints on the dynamics of scalar fields that are known to hold in flat space~\cite{deRham:2022hpx} and in AdS~\cite{Hartman:2022zik} have remained hidden in cosmic observables. We expect many of these bounds to hold in cosmology as well, but this has not been demonstrated.

\vskip 4pt
In this paper, we explored some very simple constraints on fields in both de Sitter and inflationary backgrounds using classical statistics. While these bounds will necessarily be satisfied by any realistic measurement, they imply nontrivial constraints on the dynamics of fields in the early universe. Concretely, there can be no negative (real) anomalous dimensions for heavy fields in quasi-de Sitter spacetimes. Furthermore, the bounds specify restrictive conditions on the correlation functions of fields with negative anomalous dimensions in inflationary backgrounds with a preferred time-slicing.

\vskip 4pt
Our results are similar in some ways to the bounds on the stress-tensor correlators from conformal collider physics~\cite{Hofman:2008ar,Cordova:2017zej,Kologlu:2019mfz}. While it may seem self-evident that a physical detector will only measure positive energies, the implication for the properties of CFTs are not. One might hope that a more nuanced analysis of the positivity of the Fisher information might similarly yield more insights into the physics of the early universe.

\paragraph{Acknowledgements}
We are grateful to Tim Cohen, Yi Guo, Jiashu Han, Akhil Premkumar, Kamran Salehi Vaziri and Ben Wallisch for helpful discussions. We also thank Tim Cohen for comments on the draft. 
DB is supported by a VIDI grant of the Netherlands Organization for Scientific Research (NWO) and a Yushan Professorship at National Taiwan University (NTU) funded by the Ministry of Education (MOE), Taiwan.
DG and YH are supported by the US~Department of Energy under grant~\mbox{DE-SC0009919}. CHS is supported by the Ministry of Education, Taiwan (MOE Yushan Young Scholar grant NTU-112V1039).

\newpage
% -----------------------------------------------------------------------------------------------------------------------------------------
\appendix

%%%%%%%%%%%%%%%%%

\section{Relation to In-In Correlators}\label{app:inin}

The freeze-out of superhorizon fluctuations ensures that cosmological correlators obey classical statistics and the host of inequalities which this implies. In the main text, we investigated the implications of these classical constraints on cosmological correlators. Yet, given how simple and general these results are, one might imagine that they are easy to derive directly within the in-in formalism. In this appendix, we will explore to what degree these results are already encoded in perturbation theory.

\subsubsection*{In-In Formalism}

The general definition of an in-in correlator is~\cite{Weinberg:2005vy,Weinberg:2006ac} 
\beq
\big\langle Q(t)\big\rangle=\left\langle\bar{T} \exp \left[i \int_{-\infty^+}^t \d t^{\prime}\, H_{\mathrm{int}}\left(t^{\prime}\right)\right] Q_{\mathrm{int}}(t)\, T \exp \left[-i \int_{-\infty^-}^t \d t^{\prime} \, H_{\mathrm{int}}\left(t^{\prime}\right)\right]\right\rangle ,
\eeq
where ($\bar T$) $T$ denotes (anti-)time ordering and $-\infty^\pm \equiv - \infty(1\pm i \epsilon)$.
 The operator $Q(t)$ is  defined in term of fields at a single time $t$, but is not necessarily local in space. 
 %Because it is defined at a single-time, we can introduce the interaction picture fields, $\zeta \to \zeta_{\rm int}$, as $Q[\zeta] \to Q_{\rm int}[\zeta_{\rm int}]$. 
 The $i\epsilon$ prescription of the time integral defines the interacting vacuum, just like in flat space.

\vskip 4pt
For illustrative purposes, we will consider the composite $Q \equiv \zeta^2(\x)$. Its Fourier transform is 
\beq
[\zeta^2](\k\hs)= \int \frac{\d^3 p}{(2\pi)^3} \,\zeta(\p\hs) \zeta(\k -\p\hs) \, ,
\eeq
and the two-point function is given by
\beq
\big\langle [\zeta^2](\k\hs)\hs [\zeta^2](-\k\hs)\big\rangle' \,=\, \big\langle U^{-1}(t,-\infty^+)\, [\zeta_{\rm int}^2](\k\hs) \hs [\zeta_{\rm int}^2](-\k\hs)\, U(t,-\infty^-) \big\rangle' \, ,
\label{equ:A3}
\eeq
where we have introduced
\beq
U(t_f, t_i) \equiv T \exp \left[-i \int_{t_i}^{t_f} \d t^{\prime} \,H_{\mathrm{int}}\left(t^{\prime}\right)\right] ,
\eeq
It is known to all orders in perturbation theory that $\zeta$ freezes out and becomes classical. As a result, the correlator (\ref{equ:A3}) is necessarily positive.  We would like to understand to what degree this is obvious by standard manipulations.

\subsubsection*{Positivity of In-In Correlators}

Positivity of this correlator is most apparent if we set $\epsilon = 0$, so that $-\infty^\pm \to -\infty$. The operator $U(t) \equiv U(t,-\infty)$ is then unitary and we can write (\ref{equ:A3}) as
\beq
\big\langle [\zeta^2](\k\hs)\hs [\zeta^2](-\k\hs) \big\rangle' = \big\langle U^{-1}(t) \,[\zeta_{\rm int}^2](\k\hs)\, U(t)\hs U^{-1}(t)\,[\zeta_{\rm int}^2](-\k\hs)\, U(t) \big\rangle'\,.
\eeq
Using momentum conservation and the reality condition $ \zeta^*(\k\hs)= \zeta(-\k\hs)$, we get
\beq
\big\langle [\zeta^2](\k\hs)\hs [\zeta^2](-\k\hs) \big\rangle'  = \big\langle \left| Q(\k\hs) \right|^2  \big\rangle'\,,
\eeq
where $Q(\k\hs) \equiv U^{-1}(t) \hs[\zeta_{\rm int}^2](\k\hs)\hs U(t)$  is a real operator. Furthermore, since $Q_{\rm int}$ is defined entirely in terms of free fields, this correlator is completely determined by Wick contractions and is positive. Unfortunately, most of the time integrals that define $U(t,-\infty)$ do not converge and therefore although the expression is manifestly positive, it is not useful without regulating the early-time (or high-energy) behavior.  

\vskip 4pt
The bad behavior at early times is the result of not being in the interacting vacuum. This is resolved by restoring $\epsilon \neq 0$, as is standard in QFT. Naively, we would simply repeat the above argument, however we see that it fails because $U$ is not manifestly unitary, 
\beq
U(t,-\infty(1-i \epsilon)) U^{-1}(t,-\infty(1+i \epsilon)) \neq 1 \, .
\eeq
For local non-Gaussianity, the contribution to the correlator is dominated by late times where $\epsilon=0$ can be used reliably. More generally, physics at horizon crossing does matter, as does the $i\epsilon$ precision and thus our positivity constraints are not trivial consequences of the usual in-in expressions.

\subsubsection*{Complete Set of States}

An alternative strategy is to assume that we can insert of a complete set of states, 
\beq
1 =  \sum_{n,\p} |n,\p\hs\rangle \langle n,\p \hs| \, ,
\eeq
where $n$ is some internal quantum number and $\p$ is the  momentum. %Assuming that the states and operators are all defined at the same time $t$, and that $\P(\x)= F[\zeta(\x)]$ is a local (real) operator defined in terms of $\zeta$, then
We then have\footnote{A similar argument was used in~\cite{Assassi:2012zq} to prove that $\tnl> (6 \fnl/5)^2$ for general masses, assuming that the sum is dominated by single-particle states.}
\begin{align}
\big\langle [\zeta^2](\k\hs) \hs [\zeta^2](\k^{\hskip 2pt \prime})\big\rangle  &= \sum_{n,\p}   \big\langle [\zeta^2](\k\hs) \, |n,\p\hs\rangle \langle n,\p\hs|\,
 [\zeta^2](\k^{\hskip 2pt \prime}) \big\rangle \nonumber \\
&= (2\pi)^3 \delta_D(\k +\k^{\hskip 2pt \prime}) \sum_{n}   \big\langle [\zeta^2](\k\hs)  |n,\k\hs \rangle \langle n,\k\hs|
[\zeta^2](-\k\hs) \big\rangle \nonumber \\
 &= (2\pi)^3 \delta_D(\k +\k^{\hskip 2pt \prime}) \sum_{n} \big|\langle [\zeta^2](\k\hs)  |n,\k\hs \rangle\big|^2 \,.\label{eq:complete_set}
\end{align}
This expression can be used to reproduces the positivity bounds we derived in the main text. However, given that a nonperturbative definition of the complete basis of states has not been demonstrated in context of inflation, it leaves possible loopholes that do not apply to the bounds in this paper. 

%================================================
\section{Positivity of Two-Point Functions}
\label{app:two_point}
%================================================

The fact that the two-point functions of $\O_\pm$, or their contributions to the trispectrum, are not positive when  ${\rm Re} (\Delta) \neq 3/2$ is a bit surprising. After all, these operators arise as the long-wavelength description of real scalar fields and the power spectra of real operators are positive. Specifically, if $\O(\x)$ is a real operator, then the reality condition in Fourier space implies that $\O^\dagger(\k\hs) = \O(-\k\hs)$ and therefore 
\beq
\langle \O(\k\hs) \O(-\k\hs) \rangle' = \langle |\O(\k\hs)|^2 \rangle \, ,
\eeq
which is the average of a manifestly positive quantity. In this appendix, we will explore the consequences of ${\rm Re}(\Delta) \neq 3/2$ for the operators $\O_\pm$ in consistent theories. We confirm this behavior in a simple example in dS where such operators must arise.

\subsubsection*{Two-Point Negativity}

When the scaling dimension $\Delta$ is real, it is reasonable to demand that $\O_\Delta(\x)$ is a real operator and its two-point statistics are positive. However, when $\Delta$ is complex, $\O_\Delta$ is also complex and therefore
\beq
\langle \O_\Delta(\k\hs) \O_\Delta(-\k\hs) \rangle' \neq \langle |\O_\Delta(\k\hs)|^2 \rangle \ .
\eeq
Importantly, this means that the power spectrum of $\O_\Delta$ does not have a fixed sign. 

\vskip 4pt
When $\Delta$ is complex, there always exists a second operator with dimension $\Delta^*$. It is therefore natural to assume that these operators obey $\O_{\Delta}^\dagger(\k\hs) = \O_{\Delta^*}(-\k\hs)$, in which case it should be true that
\beq
\langle \O_{\Delta}(\k\hs) \O_{\Delta^*}(-\k\hs)  \rangle' = \langle |\O_\Delta(\k\hs)|^2 \rangle \geq 0 \, .
\eeq
This is true for free principal series fields in dS. However, for complex $\Delta$, with ${\rm Re}(\Delta) \neq 3/2$, conformal invariance demands that
\beq
\langle |\O_\Delta(\k\hs)|^2 \rangle = 0 \, .
\eeq
On its own, this might appear to imply the existence of null states, but it is crucial that there is no state operator correspondence, and therefore
\beq
\langle |\O_\Delta(\k\hs)|^2 \rangle \neq \langle \Delta | \Delta \rangle \, .
\eeq
Yet, it does imply the surprising outcome that 
\begin{align}
\big\langle \left(\O_\Delta(\k\hs)+ \O_{\Delta^*}(\k\hs) \right) \left(\O_\Delta(-\k\hs)+ \O_{\Delta^*}(-\k\hs) \right) \big\rangle^\prime &= \big\langle \O_\Delta(\k\hs) \O_\Delta(-\k\hs) \big\rangle^\prime + \big\langle \O_{\Delta^*}(\k\hs) \O_{\Delta^*}(-\k\hs) \big\rangle^\prime \nonumber\\
&\propto  {\rm Re}\left(\frac{k}{aH} \right)^{2\Delta}  \, ,
\end{align}
which is {\it not} positive definite for complex $\Delta$. Hence, if $\Delta$ is complex and ${\rm Re}(\Delta) \neq 3/2$, then $\O_\Delta(\x) + \O_{\Delta^*}(\x)$ is not real, or equivalently,  $\O_{\Delta^*}^\dagger(\x) \neq \O_\Delta(\x)$.
%\beq
%\O_{\Delta^*}^\dagger(\x) \neq \O_\Delta(\x) \, .
%\eeq
Negative two-point statistics on their own is not an issue, but it does require the failure of these kinds of reality conditions.

\vskip 4pt
Of course, in the main text, we showed that ${\rm Re}(\Delta) < 3/2$ is inconsistent with the required positivity of cosmology correlators and thus are forbidden. In contrast, ${\rm Re}(\Delta) > 3/2$ also have negative two-point statistics, but these operators necessarily arise in physical theories. As a result, our bounds are not equivalent to the condition that these two-point statistics must be positive.

\subsubsection*{An Example}

For illustration, consider a principal series field producing operators $\varphi_{\pm}$ with dimensions $\Delta_\pm = 3/2 \pm i \nu$. We now look at the two-point statistics of the composite operators $[\varphi_{\pm}^2]$ with dimensions $2\Delta_\pm$. In the free theory, the power spectra of these operators are determined by Wick contractions and yield
\begin{align}
\langle [\varphi_{\pm}^2](\k\hs) [\varphi_{\pm}^2](-\k\hs) \rangle' &= \int \frac{\d^3p}{(2\pi)^3} \,p^{\pm 2i\nu} |\k-\p\hs|^{\pm 2i \nu} \nonumber \\
&= \frac{1}{2\pi^2} \frac{2^{-4\mp 2i \nu} \pi \Gamma[-\frac{3}{2}\mp  i 2\nu]\Gamma[\frac{3}{2}\pm i \nu]}{4 \Gamma[2\pi \nu]\Gamma[\mp  i \nu]^2} \,k^{3\pm 4i\nu} \, .
\end{align}
We see that the result is both nonzero and {\it not} positive. The principal series field itself, on the other hand, has a positive two-point function by virtue of the contact term
\beq
\langle \varphi_+(\k\hs) \varphi_-(-\k\hs) \rangle'  = C\,,
\eeq
which is allowed because ${\rm Re}(\Delta) = 3/2$ and thus this contact term is consistent with conformal invariance. However, there is no such term for the composite operator,
\beq
\langle [\varphi_{+}^2](\k\hs) [\varphi_{-}^2](-\k\hs) \rangle^\prime =  \int \frac{\d^3p}{(2\pi)^3} \,C^2 = 0\,,
\eeq
where we used dimensional regularization to regulate the power-law divergence. These operators are physical and yet have negative two-point statistics. We conclude that our bounds are not trivially related to positivity of the two-point statistics in de Sitter space.

\clearpage
\phantomsection
\addcontentsline{toc}{section}{References}
\small
\bibliographystyle{utphys}
\bibliography{CPRefs-V2}

\end{document}